\documentclass[11pt]{article}

\usepackage{amsmath}
\usepackage{amsfonts,amssymb}
\usepackage{epsfig}
[1999/03/29 PSNFSS v.7.2
Times font as default roman
: S Rahtz]

\newlength{\Taille}

\def\A{{\mathcal A}}
\def\B{{\mathcal B}}

\def\S{{\mathcal S}}
\def\I{{\mathcal I}}

\newcommand{\RR}{\mathbb{R}}
\newcommand{\ZZ}{\mathbb{Z}}

\newcommand{\NN}{\mathbb{N}}
\newcommand{\CC}{\mathbb{C}}

\def\<{\mathbb{E}\left(}
\def\>{\right)}

\def\qed{$\square$}

\newtheorem{definition}{Definition}
\newtheorem{theorem}{Theorem}
\newtheorem{lemma}{Lemma}

\newcommand{\flechebas}[1]{
  \settoheight{\unitlength}{\mbox{$#1$}}
  \settowidth{\Taille}{\mbox{~${\scriptstyle #1}$}}
  \addtolength{\unitlength}{4ex}
  \begin{picture}(0,1)
    \put(0,1){\vector(0,-1){1}}
    \put(0,0.5){\makebox(0,0){${\scriptstyle #1}$ \hspace{\the\Taille}}}
  \end{picture}}
\newcommand{\flechehaut}[1]{
  \settoheight{\unitlength}{\mbox{$#1$}}
  \settowidth{\Taille}{\mbox{~${\scriptstyle #1}$}}
  \addtolength{\unitlength}{4ex}
  \begin{picture}(0,1)
    \put(0,0){\vector(0,1){1}}
    \put(0,0.5){\makebox(0,0){\hspace{\the\Taille}${\scriptstyle #1}$ }}
  \end{picture}}
\newcommand{\flechedroite}[1]{
  \settowidth{\unitlength}{\mbox{$#1$}}
  \settoheight{\Taille}{\mbox{${\scriptstyle #1}$}}
  \addtolength{\Taille}{1ex}
  \addtolength{\unitlength}{4ex}
  \raisebox{0.5ex}{
  \begin{picture}(1,0)
    \put(0,0){\vector(1,0){1}}
    \put(0.5,0){\makebox(0,0){${\scriptstyle #1}$ \vspace{\the\Taille}}}
  \end{picture}}}
\newcommand{\flechegauche}[1]{
  \settowidth{\unitlength}{\mbox{$#1$}}
  \settoheight{\Taille}{\mbox{${\scriptstyle #1}$}}
  \addtolength{\Taille}{1ex}
  \addtolength{\unitlength}{4ex}
  \raisebox{0.5ex}{
  \begin{picture}(1,0)
    \put(1,0){\vector(-1,0){1}}
    \put(0.5,0){\makebox(0,0){${\scriptstyle #1}$ \vspace{\the\Taille}}}
  \end{picture}}}

\newcommand{\F}{{\cal F}}

\begin{document}

\title{Log-infinitely divisible multifractal processes}


\author{E. Bacry$^1$  and J.F. Muzy$^2$ \\
$^1$ Centre de Math\'ematiques Appliqu\'ees, Ecole Polytechnique, \\ 
91128 Palaiseau Cedex, France, \\ Email: emmanuel.bacry@polytechnique.fr \\
$^2$ CNRS, UMR 6134, Universit\'e de Corse, Grossetti, \\ 20250 Corte, France, \\
Email: muzy@univ-corse.fr \\
Submitted to Communication in Mathematical Physics}

\date{July 2, 2002}

\maketitle


\maketitle
\begin{abstract}We define a large
class of multifractal random measures and processes with 
arbitrary log-infinitely divisible exact or asymptotic scaling law.
These processes generalize within a unified framework both the recently
defined log-normal Multifractal Random Walk processes (MRW) \cite{mrwepj,mrwpre} 
and the log-Poisson ``product of cynlindrical pulses" \cite{barral}. 
Their construction involves some ``continuous  stochastic multiplication'' \cite{schmitt} from coarse to fine scales. They are obtained  as limit processes when the finest scale goes to zero. We prove the existence of these limits and 
we study their main statistical properties  
including non degeneracy, 
convergence of the moments and multifractal scaling.
\end{abstract}

\section{Introduction}
Fractal objects and the related concept of
scale-invariance, 
are now generally used in many fields of natural,
information or social sciences. 
They have been involved   
in large amount of empirical, as well as theoretical 
studies concerning a wide variety of problems.
The scale-invariance property of a stochastic
process is usually quantified by the scaling
exponents $\zeta_q$ associated with the power-law
behavior of the order $q$ moments of the ``fluctuations''
at different scales.
More precisely, for a 1D random process\footnote{We will 
exclusively consider, in this paper, real valued random functions
of a 1D continuous ``time'' variable $t$.
Though the extension to higher dimensions is rather natural, this
problem will be addressed in a forthcoming study.} $X(t)$,
let us consider the order $q$ absolute moment of the ``fluctuation''
$\delta_l X(t)$ at scale $l$:
\begin{equation}
\label{moments}
m(q,l) = \< |\delta_l X(t)|^q \> \; ,
\end{equation}
where the ``fluctuation'' process $\delta_l X(t)$ is assumed to be
stationary and $\< . \>$ stands for
the mathematical expectation.
Usually, the fluctuation $\delta_l X(t)$ is chosen
to be the increment of $X(t)$ at time $t$ and scale $l$:
\[ 
\nonumber
  \delta_l X(t) = X(t+l)-X(t)
\]
but it can be also defined as a wavelet coefficient \cite{wtmm1,wtmm2,wtmm3}.
The $\zeta_q$ exponents are defined from the
power-law scaling
\begin{equation}
\label{scaling}
m(q,l) = K_q  l^{\zeta_q}, ~\forall \; l \le T \; .
\end{equation}
When the $\zeta_q$ function is linear, i.e., $\zeta_q = qH$, the
process is referred to as a {\em monofractal} process with 
Hurst exponent $H$.
In that case the scaling can extend over an unbounded
range of scales (one can have $T = +\infty$).
Examples of monofractal processes are the so-called self-similar
processes like (fractional) Brownian motion or $\alpha$-stable
motion \cite{Taqqu}.
When the function $\zeta_q$ is non-linear, it is necessarily
a concave function and $T < +\infty$ .
In that case the process is called a {\em multifractal} process. 
Let us remark that this definition of multifractality
relies upon the scaling properties of increment absolute moments.
An alternative definition refers to the point-wise fluctuations
of the regularity properties
of sample paths (see e.g. \cite{barral,jafflevy}). 
Sometimes, one can establish an {\em exact} equivalence between these 
two definitions within the so-called {\em multifractal formalism}. 

Let us note that the scaling equation (\ref{scaling})
refers to an exact continuous scale invariance.
Weaker forms of scale invariance are often used, notably
{\em asymptotic} scale invariance that assumes that the
scaling holds only in the limit $l \rightarrow 0^+$:
\begin{equation}
\label{ascaling}
m(q,l) \sim K_q  l^{\zeta_q}, ~\mbox{when} \; \; l \rightarrow 0^+ \; .
\end{equation}
The {\em discrete} scale invariance only assumes
that the scaling holds for a discrete subset of scales
$l_n$ (with $l_n \rightarrow 0$ when $n \rightarrow +\infty$):
\begin{equation}
\label{dscaling}
m(q,l_n) = K_q  l_n^{\zeta_q} \; .
\end{equation}
The paradigm of multifractal processes 
that satisfy discrete scale invariance are
the Mandelbrot multiplicative cascades \cite{man76} or
the recently introduced wavelet cascades \cite{jmp}.
In Mandelbrot construction (the principle is the same
for wavelet cascades), $l_n = 2^{-n}$ and 
a sequence of probability measures
$M_{l_n}(dt)$ is built recursively.
$M_{l_n}(dt)$ is uniform on dyadic intervals $\I_{n,k} = ]k2^{-n},(k+1)2^{-n}]$
and is obtained from $M_{l_{n-1}}(dt)$ using the cascading rule:
\begin{equation}
\label{dc}
  M_{l_n}(dt) = W_{n,k} M_{l_{n-1}}(dt), \; \mbox{for} \; t \in \I_{n,k} 
\end{equation} 
where the weights $W_{n,k}$ are i.i.d. positive random variables
such that $\< W_{n,k} \> = 1$.
The convergence and regularity properties of such construction
have been studied extensively \cite{kp,dl,guiv,molchan}
(from a general point of view, convergence of multiplicative
constructions to singular measures have been studied in \cite{kcm}).
Despite the fact that multiplicative discrete cascades
have been widely used as reference models in many applications,
they possess many drawbacks related to their discrete scale
invariance, mainly they involve a particular scale ratio (e.g. $\lambda = 2$)
and they do not possess stationary fluctuations (this comes from the fact that they
are constructed on a dyadic tree structure).

The purpose of this paper is to define a new class of continuous time 
stochastic processes with stationary fluctuations and 
that are multifractal in the sense that they verify exact
or asymptotic continuous 
scaling (Eqs. (\ref{scaling}) or (\ref{ascaling})) with 
a non linear $\zeta_q$ spectrum.
Though continuous time multifractal processes
with continuous scale invariance are obviously appealing 
from both fundamental and modeling aspects, 
until very recently such processes were lacking.
From our knowledge, only the  recent works by
Bacry et al. \cite{mrwepj,mrwpre} and Barral and Mandelbrot \cite{barral} 
refer to a precise construction of multifractal continuous scale
invariant processes.
Bacry et al. have built the so-called Multifractal Random Walk (MRW) processes
as continuous time limit processes based on discrete time random walks
with stochastic log-normal variance. 
Independently, Barral and Mandelbrot \cite{barral} have proposed a new class
of stationary multifractal measures. Their construction is
based on distributing, in a half-plane, Poisson points associated with i.i.d.
random weights and then taking a product of these weights over conical domains.
The construction that we propose in this paper generalizes 
these two approaches to
an unified framework involving general infinitely divisible laws.
Within this new framework, the constructions of \cite{mrwpre}
and \cite{barral} are particular cases 
respectively associated with normal
and compound Poisson distributions. 
The main principles of this contruction are similar to those of
Barral and Mandelbrot construction \cite{barral} and 
have been previously described in \cite{delour}.
They rely upon an idea advanced in a recent work by
Schmitt and Marsan \cite{schmitt}. These authors have shown that
a ``continuous limit" of the discrete cascade equation (\ref{dc})
can be interpreted as the exponential of a stochastic integral
of an infinitely divisible 2D noise over a suitably chosen conical
domain. We exploit this idea in order to construct 
random measures and random processes that satisfy either the exact
scaling (\ref{scaling}) or the asymptotic scale (\ref{ascaling}) 
with a $\zeta_q$ function
that can be associated with an almost 
arbitrary infinitely divisible law. More specifically, we show that
this new class of processes satisfies a {\em continuous cascade}
equation: the 
fluctuation process $\{\delta_{\lambda l} X(t)\}_t$ 
at a scale $\lambda l$ (where $l$ is an arbitrary scale
smaller than the large scale $T$ and $\lambda < 1$) 
is obtained from the
fluctuation process $\{\delta_{l} X(t)\}_t$ at the larger scale $l$
through the simple ``cascading'' rule
\begin{equation}
\label{cascade}
\{\delta_{\lambda l} X(t)\}_t \operatornamewithlimits{=}^{law} W_\lambda \{\delta_l X(t) \}_t, 
\end{equation}
where $\ln(W_\lambda)$ 
is an arbitrary infinitely divisible random variable 
independent of $\{ \delta_l X(t) \}_t$ and which
law depends only on $\lambda$.
Let us mention that, in the companion paper \cite{premrw}, 
the random processes introduced in this paper
have been also studied with less care for rigor but with many intuitive 
arguments, numerous examples, methods for simulations and
a discussion of possible applications .

The paper is organized as follows.  
In section \ref{fd} we introduce our main notations
and some well known results about independently scattered random measures \cite{rosinski}. In section \ref{MRMMdef} we define a new class of
stationary random measures: the Multifractal Random Measures (MRM). Section \ref{MRMres}  states the main results on these measures (mainly, non degeneracy, convergence of the moments and exact or asymptotic multifractal scaling). These results are proved all along section  \ref{MRMproof}.
In section \ref{MRWdef}, MRM and brownian motion are used
to build a family
of continuous time multifractal random processes : the log-infinitely divisible Multifractal Random Walks (MRW).The main results concerning MRW are given (and proved) in section \ref{MRWres}. Particular cases of MRM and MRW along with connected approaches for building multifractal stochastic measures or processes
are discussed in section \ref{examples}.
Conclusions and prospects for future research are 
reported in section \ref{conclusion}.
Some technical proofs are detailed in Appendices.

\section{Basic Notations - Independently scattered random measures}
\label{fd}
\subsection{Basic notations}
Hereafter the symbol $\< X \>$ will stand for the 
mathematical expectation of the random variable $X$. We will always omit the reference to the randomness parameter. If $X$ is a function of 
time $t$,  $X(t)$ will denote the random variable at fixed time $t$ whereas
$\{ X(t) \}_t$ will denote the whole process. 
The equality $\operatornamewithlimits{=}^{law}$
will denote the equality of finite
dimensional distributions.  
Moreover, the abbreviation {\em a.s} (resp. {\em m.s.}) 
stands for {\em almost surely} (resp. {\em mean square}).

Let us define the measure space $(\S^+,\mu)$ as follows.
$S^+$ is the space-scale half-plane 
\[ S^+ =  \{(t,l),~t\in \RR,~l \in \RR^{+\ast}\} \] 
with which one can associate the measure
\[ \mu(dt,dl) = l^{-2} dt dl \; .\]
This measure is the (left-) Haar measure of
the translation dilation group acting on $\S^+$.

\subsection{Infinitely divisible random variables and 
independently scattered random measures}
Let us recall \cite{Feller} that a random variable 
$X$ is {\em infinitely divisible}
iff, for all $n \in \NN^{\ast}$ 
\[ 
\nonumber
  X \operatornamewithlimits{=}^{law} \sum_{i=1}^n X_{n,i}
\]
where $X_{n,i}$ are i.i.d. random variables.
Infinitely divisible random variables
are intimately related to Levy Processes \cite{Feller,Bertoin} 
that are stochastic processes with independent increments.
The characteristic function of an infinitely divisible
random variable $X$, can be written as
\[ \nonumber
  \< e^{iqX} \> = e^{\varphi(q)}
\]
where $\varphi(q)$ is characterized by the 
the celebrated L\'evy-Khintchine formula \cite{Feller,Bertoin}:
\begin{equation}
\label{LK}
  \varphi(q) = imq +\int \frac{e^{iqx}-1-iq \sin x}{x^2} \nu(dx).
\end{equation}
where $\nu(dx)$ is the so-called {\em Levy measure} and
satisfies $\int_{-\infty}^{-y} \nu(dx)/x^2 < \infty $ and 
$\int_{y}^{\infty} \nu(dx)/x^2 < \infty $ for all $y > 0$.

Following \cite{rosinski}, one can introduce 
an independently scattered infinitely divisible 
random measure  
$P$ distributed on the half-plane
$\S^+$.
``$P$ is independently scattered'' means that, 
for every sequence of disjoint sets ${\A_n}$ of $\S^+$,
$\{P(\A_n)\}_n$ are independent random variables and 
\[ \nonumber
  P \left(\cup_{n=1}^{\infty} \A_n \right)= \sum_{n=1}^{\infty} P(\A_n)\;, \; \; \mbox{a.s.}
\]
provided $\cup_{n=1}^{\infty} \A_n \subset \S^+$.
$P$ is said to be infinitely divisible on $(S^+,\mu(dt,dl))$
associated with
the Levy measure $\nu(dx)$ if
for any $\mu$-measurable set $\A$, 
$P(\A)$ is an infinitely divisible random variable
which characteristic function is 
\begin{equation}
\label{meerde}
 \< e^{iqP(\A)} \> = e^{\varphi(q)\mu(\A)},
\end{equation}

Let us notice that one can build a convex function 
$\psi(q)$ ($q \in \RR^+$) such that, 
for all (non empty) subset $\A$ of $\S^+$, 
\begin{itemize}
\item[$\bullet$] $\psi(q) = +\infty$,~ if $\<e^{qP(\A)}\> = +\infty$,
\item[$\bullet$] $\<e^{qP(\A)}\> = e^{\psi(q)\mu(\A)}$,~ otherwise.
\end{itemize}
Moreover, if we define
\begin{equation}
\label{qc}
 q_c = \max_q \{q \ge 0, \; ~\psi(q) < +\infty\},
\end{equation}
it is then clear that $\forall q \in [0,q_c[, ~\psi(q) \neq +\infty$.
Let us note that, one can extend the definition of $\varphi$ 
so that it is a continuous function of a complex variable such that
\[ \nonumber
  \psi(q) = \varphi(-iq), ~\forall q \in \{z \in \CC,~0\le Re(z) < q_c\}.
\]
Moreover, in the case, there exists $\epsilon > 0$ such that 
$\<e^{-\epsilon P(\A)}\> < +\infty$, 
for a non empty subset $\A$ of $\S^+$, $\varphi$ 
can be chosen to be analytical in the strip 
$\{z \in \CC,~-\epsilon< Re(z) < q_c\}$.

\section{Definitions of Multifractal Random Measures (MRM)}
\label{MRMMdef}
\subsection{Defining the generic MRM measure $M(dt)$}\label{MRMdefM}
Let $P$ be 
an infinitely divisible independently scattered random measure 
on $(S^+,dtdl/l^2)$ as defined in the previous section,
associated with
the Levy measure $\nu(dx)$ and such that $q_c > 1$, i.e.,
\begin{equation}
\label{toto}
\exists \; \epsilon > 0,~~\psi(1+\epsilon) < +\infty.
\end{equation}
\begin{definition} {\bf (Filtration $\F_l$)} \\
Let $\Omega$ be the probability space on which $P$ is defined. 
$\F_l$ is the filtration 
of $\Omega$ defined by $ \F_l = \Sigma \{ P(dt,dl'),~l'\ge l \}$. \end{definition}
The construction of $M$ involves cone-like subsets of $\S^+$. 
These cone-like subsets $\A_l(t)$ are defined using a boundary function $f(l)$.
More precisely,
\begin{definition} {\bf (the $\A_l(t)$ set and the $f(l)$ function)} \\
The subset $\A_l(t)$ of $\S^+$ is defined by
\begin{equation}\label{Al}
\A_{l}(t) = \{(t',l'), ~ l'\ge l,~-f(l')/2 < t'-t\le f(l')/2\}.
\end{equation}
where $f(l)$ is a positive function of $l$ such that 
\[
  \int_l^{+\infty} f(s)/s^2 ds  <  +\infty.
\]
\[
  \exists \; L > 0 \;,  f(l)  =  l \; \mbox{for} \; l < L \; .
\]
\end{definition}
Let us note that $\mu(\A_l)  = \int_l^{+\infty} f(s)/s^2 ds < +\infty$.
It is convenient to represent the function $f(l)$ as:
\begin{equation}
\label{fspec}
f(l) = f^{(e)}(l) + g(l),
\end{equation}
where $f^{(e)}$ is defined by (\ref{fexact}) and $g$ satisfies 
\[ \nonumber
\exists \; L>0,~\forall l < L,~~ g(l) = 0.
\]
\begin{definition} {\bf ($\omega_l(t)$ process)} \label{defomega} \\
The process $\omega_l(t)$ is defined as
\begin{equation}
\label{omegal}
\omega_l(t) = P\left(\A_l(t)\right),
\end{equation}
\end{definition}

\begin{definition} {\bf ($M_l(dt)$ measure)} \\
\label{mldef}
For $l > 0$, we define 
\begin{equation}
\label{mldefeq}
M_l(dt) = e^{\omega_l(t)} dt,
\end{equation}
in the sense that for any Lebesgue measurable set $I$, one has
\[ \nonumber
M_l(I) = \int_I e^{\omega_l(t)} dt.
\]
\end{definition}
Let us note that (\ref{mldefeq}) makes sense because $\omega_l(t)$ corresponds almost surely to a 
right continuous and left-hand limited (cadlag) 
function of $t$. Indeed, it can be expressed as 
\[ \nonumber
\omega_l(t) = X_l(t)-Y_l(t) +Z_l,
\]
with 
\[ \nonumber
X_l(t) = P(\{t',l')~l'\ge l,~f(l')/2\le t' \le t+f(l')/2\}),
\]\[ \nonumber
Y_l(t) = P(\{t',l')~l'\ge l,~-f(l')/2\le t' \le t-f(l')/2\}),
\]
and
\[ \nonumber
Z_l = P(\{t',l')~l'\ge l,~-f(l')/2\le t' \le f(l')/2\}.
\]
One can easily check that $\{X_l\}_{t}$ and $\{Y_l(t)\}_{t}$ are
Levy processes (with $X_l(0) =Y_l(0) = 0$). Since Levy processes are well known to have almost surely 
cadlag versions and since $Z_l$ does not depend on $t$, then
$\omega_l(t)$ is also cadlag. \qed

\begin{definition} {\bf ($M$ measure)} \\
\label{mdef}
The MRM measure $M(dt)$ is  defined as the limit measure (when it exists)
\[ \nonumber
M(dt) = \lim_{l \rightarrow 0^+} M_l(dt).
\]where $M_l(dt)$ is defined in definition \ref{mldef}.
\end{definition}
Since a simple change in the mean of the stochastic measure $P$ would lead to the same MRM measure up to a multiplying factor, we will assume, without loss of generality that $\psi$ satisfies
\[ \nonumber
\psi(1) = 0.
\]
\subsection{Defining the exact scaling MRM measure $M^{(e)}(dt)$}
\label{MRMEdef}
In the previous section, for any choice of $f(l)$, 
one can define an MRM $M(dt)$. In the following, 
we will prove that  $M([0,t])$ satisfies 
the asymptotic scaling (\ref{ascaling}). 
From a fundamental point of view
but also in most applications, it is interesting
to have a model where
the scale invariance property (\ref{scaling}) 
is ``measured'' on a whole range of scales and
not only as an asymptotic property.
One thus needs to build multifractal processes/measures 
that satisfy the exact scaling relation (\ref{scaling}) 
on a whole range of scales $l \in ]0,T]$, 
where $T$ is an arbitrary large scale. 
The exact scaling property  can be obtained by picking 
up a particular shape for the 
set $\A_l$, i.e., by choosing the appropriate function $f(l)$.
As we will see, only the particular choice $f(l)=f^{(e)}(l)$
(i.e. $g(l) = 0$ in (\ref{fspec})) with
\begin{equation}
\label{fexact}
f^{(e)}(l) = \left\{
\begin{array}{lll}
& l, & \forall l \le T \\
& T,  & \forall l>T 
\end{array}
\right.
\end{equation}
will lead to a MRM measure with exact scaling (\ref{scaling}). The MRM measure associated to this particular choice will be referred to as $M^{(e)}(dt)$, i.e.,
\begin{definition} {\bf ($M^{(e)}$ measure)} \\
The MRM measure $M^{(e)}(dt)$ is  defined as the limit measure (when it exists)
\[ \nonumber
M^{(e)}(dt) = \lim_{l \rightarrow 0^+} M^{(e)}_l(dt).
\]where $M^{(e)}_l(dt)$ is equal to $M_l(dt)$ (definition  \ref{mldef}) for the particular choice $f = f^{(e)}$ (Eq. (\ref{fexact})).
\end{definition}

\subsection{Numerical simulation - Discrete construction of an MRM}
\label{MRMDdef}
In the case the Levy measure $\nu(dx)$ verifies $\int \nu(dx)/x^2 < +\infty$, a realization of the measure $P(dt,dl)$ is made of isolated weighted dirac distribution in the $\S^+$ half-plane (the construction then basically reduces to the one of Barral and Mandelbrot \cite{barral}).
Thus the process $M_l(([0,t[) = \int_0^t e^{\omega_l(t)} dt$ is a jump process that can be simulated with no approximation. This gives a way of simulating a process which is arbitrary close (by choosing $l$ close to 0) to the limit process.

However, if $\int \nu(dx)/x^2 = +\infty$ (e.g., $\nu(dx)$ has a gaussian  component), this is no longer the case.
Thus, one has to build another sequence of stochastic measures $\tilde M_l(dt)$ that converges towards $M(dt)$ and that can be easily simulated.  The  purpose of the following construction is to build such a sequence.

We choose $\tilde M_l(dt)$ to be uniform on each interval of the form $[kl,(k+1)l[$, $\forall k \in \NN$ and with density $e^{\omega_l(kl)}$. Thus, for any $t > 0$ such that $t = nl$ with $n \in \NN^*$, one gets
\begin{equation}\label{tmldef}
\tilde M_l([0,t[) = \sum_{k=0}^{n-1} e^{\omega_l(kl)} l.
\end{equation}
We will restrict ourselves to $l = l_n = 2^{-n}$. We then define the discretized MRM $\tilde M$ :
\begin{definition} {\bf ($\tilde M$ measure)} \\
The discretized MRM measure $\tilde M(dt)$ is  defined as the limit measure (when it exists)
\begin{equation}\label{tmdef}
{\tilde M}(dt) = \lim_{n \rightarrow +\infty} {\tilde M}_{l_n}(dt),
\end{equation}
where $\tilde M_{l_n}(dt)$ is defined by (\ref{tmldef}) with $l_n = 2^{-n}$.
\end{definition}

\section{Main results on MRM}
\label{MRMres}
In this section, we state the main theorems concerning MRM measures. 
All these theorems are proved all along the next section.
Most of the results will first be proved for $M = M^{(e)}$.
The generalization to any measure $M$ will then be made using
the exact scaling properties of $M^{(e)}$ (theorem \ref{TMRMescaling}).
We introduce the following definition
\begin{definition} {\bf (scaling exponents $\zeta_q$)} \\
\begin{equation}
\label{zetaq}
\forall q>0,~~~\zeta_q = q-\psi(q).
\end{equation}
\end{definition}
Note that because $\psi$ is convex, $\zeta_q$ is a concave function.
As we will see in the following theorems, the so-defined exponents $\zeta_q$ do correspond to the multifractal scaling exponents in the sense of (\ref{scaling}) or (\ref{ascaling}).
\begin{theorem} {\bf (Existence of the limit MRM measure $M(dt)$)} \\
\label{TMRMexist}\label{TFIRST}
There exists a stochastic measure $M(dt)$ such that 
\begin{itemize}
\item[(i)] almost surely $M_l(dt)$ converges weakly towards $M(dt)$, for $l\rightarrow 0^+$, 
\item[(ii)] $\forall t \in \RR$, $M(\{t\}) = 0$,
\item[(iii)] for any bounded set $K$ of $\RR$, $M(K)<+\infty$ and $\<M(K)\> \le
|K|$.
\end{itemize}
\end{theorem}
\begin{theorem}
\label{TMRMdeg} {\bf (Non degeneracy of $M(dt)$)} \\
(H)~$\exists \; \epsilon > 0,~\zeta_{1+\epsilon} > 1$ \\
if (H) holds then $\<M([0,t])\> = t.$
\end{theorem}

\begin{theorem}
\label{TMRMmom} {\bf (Moments of positive orders of $M(dt)$)} \\
Let $q > 0$ then
\begin{itemize}
\item[(i)] $ \zeta_{q} > 1 \Longrightarrow  \<M([0,t])^{q}\>< +\infty$. 
\item[(ii)] if (H) then $\<M([0,t])^q\> < +\infty \Longrightarrow \zeta_q \ge 1$.\end{itemize}
\end{theorem}

\begin{theorem}
\label{TMRMescaling} {\bf (Exact scaling of $M^{(e)}(dt)$)} \\
\[ \nonumber
\{M^{(e)}([0,\lambda t])\}_t ~\operatornamewithlimits{=}^{{law}} W_{\lambda}  \{M^{(e)}([0, t])\}_t,~~~\forall \; \lambda \in \; ]0,1[ \; \mbox{and} \; \; t \leq T \; ,
\]
with $W_{\lambda} = \lambda  e^{\Omega_{\lambda}}$
where $\Omega_\lambda$ is an infinitely divisible 
random variable (independent of $\{M_{l}([0, t])\}_t$)  which characteristic function is
\[ \nonumber
\< e^{iq\Omega_{\lambda}}\> = \lambda^{-\varphi(q)}.
\]
If  $\zeta_q = -\infty$, then $\<M^{(e)}([0,t])^{q}\> =+\infty$ and otherwise
\begin{equation}
\label{mescaling}
\<M^{(e)}([0,t])^{q}\> = \left(\frac{t}{T}\right)^{\zeta_{q}} \<M^{(e)}[0,T])^{q}\>, ~~\forall t \le T.
\end{equation}
\end{theorem}
Let us note that this theorem shows that $M^{(e)}$ is a ``continuous
cascade'' as defined in (\ref{cascade}), i.e., it satisfies
a continuous version of the discrete multiplicative cascade recurrence
(\ref{dc}).

\begin{theorem}
\label{TMRMascaling} {\bf (Asymptotic scaling of $M(dt)$)} \\
If (H) holds, then for $q > 0$ such that $\<M([0,t])^{q}\> < +\infty$ and $\exists \; \epsilon> 0$, 
$\zeta_{q+\epsilon} \neq -\infty$,
then
\[ \nonumber
\<M([0,t])^{q}\>  ~\operatornamewithlimits{\sim}_{{t\rightarrow 0^+}} ~ \left(\frac{t}{T}\right)^{\zeta_{q}} \<M([0,T])^{q}\>.
\]\end{theorem}

\begin{theorem} {\bf (Link between $\tilde M(dt)$ and $M(dt)$)} \\
\label{Ttildem}\label{TLAST}
If there exists $\epsilon >0$ such that $\zeta_{2+\epsilon} > 1$ then
one has
\[ \nonumber
\tilde M_{l_n}(dt) ~\operatornamewithlimits{\rightarrow}^{m.s.}   M(dt),
\]where the limit is taken in the mean square sense.
\end{theorem}
\section{Proofs of theorems \ref{TFIRST} through \ref{TLAST}}
\label{MRMproof}
\subsection{Existence of the limit MRM measure $M(dt)$ - Proof of theorem \ref{TMRMexist}}
Since $\psi(1) = 0$, one has $\<e^{\omega_l(t)}\> = 1$. 
It is then easy to prove that for all Lebesgue measurable 
set $I$, 
the sequence $\{M_l(I)\}_l$ is a left 
continuous positive 
martingale with respect to  $\F_l$. 
From the general theory of \cite{kchi},
one gets theorem \ref{TMRMexist}.

\subsection{Computation of the characteristic function of  $\omega_l(t)$}
Let $q\in \NN^*$, $\vec{t}_q = (t_1,t_2,\ldots,t_q)$ with $t_1 \le t_2 \le \ldots \le t_n$ and 
$\vec{p}_q = (p_1,p_2,\ldots,p_q)$. 
The characteristic function of the vector 
$\{\omega_l(t_m)\}_{1\le m\le q}$ is defined by
\[ \nonumber 
Q_l(\vec{t}_q,i\vec{p}_q) = \< e^{\sum_{m=1}^q i p_m P(\A_{l}(t_m))}\>.
\]
Relation (\ref{meerde}) allows us to get an analytical expression for
quantities of the form
$\< e^{\sum_{m=1}^q a_m P(\B_m)}\>$ where $\{\B_m\}_m$ would be disjoint
sets in $\S^+$ and $a_m$ arbitrary complex numbers. However the
$\{\A_{l}(t_m)\}_m$ have no reason to be disjoint sets. We need to find a
decomposition of $\{\A_{l}(t_m)\}_m$ onto disjoint domains. This is
naturally done by considering the different intersections between these
sets. Let us define
\[ \nonumber
\A_l(t,t') = \A_l(t)\cap\A_l(t').
\]
In Appendix \ref{Achar}, we prove that 
\begin{lemma}
\label{lmchar}
{\bf (Characteristic function of $\omega_l(t)$)} \\
Let $q\in \NN^*$, $\vec{t}_{q} = (t_1,t_2,\ldots,t_q)$ with $t_1 \le t_2 \le \ldots \le t_n$ and 
$\vec{p}_{q} = (p_1,p_2,\ldots,p_q)$. 
The characteristic function of the 
vector $\{\omega_l(t_m)\}_{1\le m\le q}$ is
\begin{equation} 
\label{char}
\< e^{\sum_{m=1}^q i p_m P(\A_{l}(t_m))}\>  = e^{\sum_{j=1}^q \sum_{k=1}^{j}
\alpha(j,k) \rho_l(t_k-t_j)},
\end{equation}
where
\[ \nonumber
\rho_l(t) =  \mu(\A_l(0,t)),
\]
and
\[ \nonumber 
\alpha(j,k) = \varphi(r_{k,j})+\varphi(r_{k+1,j-1})-\varphi(r_{k,j-1})-\varphi(r_{k+1,j}),
\]
and
\[ \nonumber
r_{k,j} = 
\left\{
\begin{array}{lll}
& \sum_{m=k}^j p_m, & {\mbox{for}}~k \le j \\
& 0,  & {\mbox{for}} ~k > j 
\end{array}
\right.
\]
Moreover
\begin{equation}
\label{rk}
\sum_{j=1}^q \sum_{k=1}^{j}\alpha(j,k) = \varphi \left( \sum_{k=1}^q p_k \right).
\end{equation}
\end{lemma}

\subsection{Exact scaling of $M^{(e)}(dt)$ - Proof of theorem \ref{TMRMescaling}}
Let us assume that the family of processes $M_l([0,t])$ satisfy (\ref{cascade}), i.e., for $l$ small enough,  and for all $\lambda \in ]0,1[$,
\[ \nonumber
\{M_{\lambda l} ([0,\lambda t])\}_t \operatornamewithlimits{=}^{{law}}  W_{\lambda}  {M_{l} ([0,t])}_t,
\]
where $W_\lambda$ is independent of $\{M_{l} ([0,t])\}_t$.
By definition of $M_l(dt)$, it gives
\[ \nonumber
\{\int_0^{\lambda t} e^{\omega_{\lambda l}(u)}du\}_t \operatornamewithlimits{=}^{{law}}  W_{\lambda}  \{\int_0^t e^{\omega_{l}(u)}du\}_t.
\]
This last relation will hold if  the processes $\omega_l(t)$ satisfy the scaling property
\begin{equation}
\label{scalingomega1}
\{\omega_{\lambda l}(\lambda t)\}_t \operatornamewithlimits{=}^{{law}}  \Omega_{\lambda} + \{\omega_{l}(t)\}_t,
\end{equation}
where $\Omega_\lambda$ and $W_\lambda$ are linked by the relation $W_\lambda = \lambda e^{\Omega_\lambda}$.
Indeed, one would have
\begin{eqnarray}\nonumber
\{M_{\lambda l} ([0,\lambda t])\}_t  &= &\{\int_0^{\lambda t} e^{\omega_{\lambda l}(u)}du\}_t 
=  \lambda \{\int_0^{t} e^{\omega_{\lambda l}(\lambda u)}du\}_t \\
\nonumber
&\operatornamewithlimits{=}^{law}&   \lambda e^{\Omega_{\lambda}} \{\int_0^{t} e^{\omega_{ l}(u)}du\}_t =  \lambda e^{\Omega_{\lambda}} \{M_{l} ([0,t])\}_t.
\end{eqnarray}
Equation (\ref{scalingomega1}) translates easily on the characteristic function found in lemma \ref{lmchar}. This gives the following lemma
\begin{lemma} ({\bf Exact scaling of $M_l(dt)$}) \\
\label{lmexact}
Let fix $T\in\RR^{+*}$.
If $\rho_l(t) = \mu(\A_l(0,t)) = \mu(\A_l(0) \cap \A_l(t)),$ satisfies the scaling relation
\begin{equation}
\label{rhoscal}
\rho_{\lambda l}(\lambda t) = -\log{\lambda} +\rho_l(t),~~~\forall l \in ]0,T],~\forall \lambda \in ]0,1[,
\end{equation}
then one has
\begin{equation}
\label{scalingomega}
\{M_{\lambda l}([0,\lambda t])\}_t \operatornamewithlimits{=}^{{law}} \lambda  e^{\Omega_{\lambda}}  \{M_{l}([0, t])\}_t,~~~\forall l \in ]0,T],~\forall \lambda \in ]0,1[,
\end{equation}
where $\Omega_\lambda$ is an infinitely divisible random variable (independent of $\{M^{(e)}([0, t])\}_t$) which characteristic function is
\[ \nonumber
\< e^{iq\Omega_{\lambda}}\> = \lambda^{-\varphi(q)}.
\]
Moreover,  (\ref{rhoscal}) is satisfied in the particular case where $\A_l(t)$ is defined by (\ref{Al}) with $f(l)=f^{(e)}(l) $, where $f^{(e)}$ is defined by (\ref{fexact}).
\end{lemma}
(\ref{scalingomega}) is a direct consequence of the previous discussion and
lemma \ref{lmchar}. For the last assertion, one has to 
compute $\rho_{l}(t) = \rho_{l}^{(e)}(t) $ in the particular case 
$f(l) = f^{(e)}(l)$. A direct computation shows that
\begin{equation}
\label{rhoexact}
\rho_{l}^{(e)}(t) = 
\left\{
\begin{array}{ll} 
  \ln \left(\frac{T}{l} \right)+1-\frac{t}{l} & \mbox{if}~t \leq l \\
\ln \left(\frac{T}{t} \right) & \mbox{if}~T \ge t\ge l \\
0 & \mbox{if}~ t > T
\end{array}
\right.,
\end{equation} 
which satisfies (\ref{rhoscal}). \qed \\
By taking the limit $l \rightarrow 0^+$ one gets theorem \ref{TMRMescaling}.
\subsection{Moments of positive orders of $M^{(e)}(dt)$ - Proof of theorem \ref{TMRMmom} in the case $M=M^{(e)}$}
We are now ready to give conditions for existence of the moments of $M^{(e)}(dt)$.
\begin{lemma}
\label{LMRMmom}
{\bf (Moments of positive orders of $M^{(e)}(dt)$)} \\
If $q > 0$ then
\begin{itemize}
\item[(i)] $ \zeta_{q} > 1 \Longrightarrow  \<M^{(e)}([0,t])^{q}\>< +\infty$ and $\sup_l \<M^{(e)}_l([0,t])^{q}\> < +\infty$. 
\item[(ii)] if $M^{(e)} \neq 0$ then $\<M^{(e)}([0,t])^q\> < +\infty \Longrightarrow \zeta_q \ge 1$.\end{itemize}
\end{lemma}
This lemma is proved in appendix \ref{Amom}. It gives theorem \ref{TMRMmom} in the particular case $M=M^{(e)}$.

\subsection{Non degeneracy of $M^{(e)}(dt)$ - Proof of theorem \ref{TMRMdeg} 
for  the case $M=M^{(e)}$} \label{nondeg} 
The theorem \ref{TMRMdeg} for  the case $M=M^{(e)}$ is a direct consequence
of the lemma \ref{LMRMmom} (i)  using a dominated convergence argument.

\subsection{Extension of the results on $M^{(e)}(dt)$ to  $M(dt)$}
\begin{lemma} {\bf (Degeneracy, asymptotic scaling and moments of positive orders of $M(dt)$) } \\
\label{lmascaling}Let $M(dt)$ be the MRM measure as defined in definitions \ref{mldef} and \ref{mdef} with $f$ satisfying (\ref{fspec}).
Then, one has
\begin{itemize}
\item[(i)]
$M^{(e)}(dt)  \operatornamewithlimits{{=}}^{{a.s.}} 0 \Longleftrightarrow M(dt) \operatornamewithlimits{=}^{{a.s.}} 0$,
\end{itemize}
Let $q \in \RR^{+*}$. If $\exists \; \epsilon > 0$, $\zeta_{1+\epsilon} > 1$, 
and consequently (according to section \ref{nondeg}), $M^{(e)}(dt)$ and (accorging to (i)) $M(dt)$ are not degenerated. One has
\begin{itemize}
\item[(ii)] almost surely
$\forall t\ge 0,~~ M([0,\lambda t]) \sim  X M^{(e)}([0,\lambda t])$, when $\lambda \rightarrow 0^+$,
where $X$ is a random variable independent of $t$ and $\lambda$.
\item[(iii)] 
$\<M^{(e)}([0,t])^q\>  < +\infty \Longleftrightarrow \<M([0,t])^q \>  < +\infty$. \\Moreover if one this assertion holds, one has
$\sup_l\<M_l([0,t])^q\> < +\infty$.
\item[(iv)] if $\exists \; \eta>0$, $\zeta(q+\epsilon) \neq -\infty$ and $\<M^q([0,t])\> < +\infty$ then
\[ \nonumber
\<M([0,t])^q\> \operatornamewithlimits{\sim}_{{t\rightarrow 0^+}}  \left(\frac{t}{T}\right)^{\zeta_q} \<M([0,T])^{q}\>.
\]
\end{itemize}
\end{lemma}
The proof of this lemma can be found in appendix \ref{Aascaling}. 

\subsection{Proofs of theorems \ref{TMRMdeg},  \ref{TMRMmom},  and \ref{TMRMascaling}}
Using lemma \ref{LMRMmom} (i) along with lemma \ref{lmascaling} (iii), one gets
theorem \ref{TMRMdeg}. Again, lemma \ref{LMRMmom} along with lemma  \ref{lmascaling} (iii) give theorem
\ref{TMRMmom}. Lemma \ref{lmascaling} (iii) and (iv) give theorem \ref{TMRMascaling}.

\subsection{Theorem on  $\tilde M$ - Proof of theorem \ref{Ttildem}}Theorem \ref{Ttildem} is proved in appendix \ref{a1}.

\section{Defining a MRW process using a gaussian white noise}\label{MRWdef}
Let us note that attempts to define a MRW process using  a fractional gaussian noise are reported in \cite{premrw}. In this paper we will address only the case of gaussian white noise.
We define a stochastic process which is not a strictly increasing process.
The simplest approach to build such a process simply consists in subordinating a Brownian motion $B(t)$ using the MRM $M(t)$.
The subordination of a Brownian process with a non decreasing
process has been introduced by Mandelbrot and Taylor \cite{aMan67}
and is the subject of an extensive literature in mathematical finance.
Multifractal subordinators have been considered
by Mandelbrot and co-workers \cite{aMan99} and widely used
to build multifractal processes.
Let us define the process $X^{(s)}(t)$ as:
\begin{definition} {\bf (Subordinated MRW)} \\
Let $B(t)$ a brownian motion (with $\<B(1)^2\> = 1$) and  $M(dt)$ a non degenerated MRM measure which is independent of $B(t)$. The subordinated MRW process is the process defined by
\[ \nonumber
X^{(s)}(t) = B(M([0,t])).
\]\end{definition}
In the case the MRM measure that it used is $M^{(e)}$, we note
\begin{definition} {\bf ($X^{(e)}(t)$ process)}
\[ \nonumber
X^{(e)}(t) = B(M^{(e)}([0,t])).
\]\end{definition}
An alternative construction would consist in a stochastic integration using a Wiener noise $dW(u)$.
\begin{definition} {\bf (Alternative subordinated MRW)}
Let $dW(u)$ be a Wiener process of variance 1 and $\omega_l(u)$ the previously introduced process (Eq. (\ref{omegal})) such that it is independent of $dW(u)$.
The MRW process is defined as the limit (when it exists)
\[ \nonumber
X(t) = \lim_{l \rightarrow 0^+} X_l(t),
\]
where
\begin{equation}
\label{defmrw}
X_l(t) = \int_0^t e^{\omega_l(u)/2} dW(u),
\end{equation}
\end{definition}
Let us note that the process $X_l(t)$ is well defined since $\int_0^t \< (e^{\omega_l(u)/2})^2\> = \int_0^t \< e^{\omega_l(u)} \>  = t$.
As we will see in the next section, 
these two constructions lead to the same process.

As explained in section \ref{MRMDdef}, for simulation  purposes, one needs to define a discrete time process that converges towards the MRW. This is done naturally using the discretized MRM measure $\tilde M$ introduced in section 
\ref{MRMDdef}.
\begin{definition}
Let $\{w[k]\}_{k\in\ZZ}$ be a gaussian white noise of variance 1. Let $l_n = 2^{-n}$.
We define the discretized MRW $\tilde X_(t)$ as the limit process (when it exists)\[ \nonumber
\tilde X(t) = \lim_{l \rightarrow 0^+} \tilde X_{l_n}(t),
\]
where
\[ \nonumber
\tilde X_{l_n}(t) = \int_0^t \sqrt{\tilde M_{l_n}(du)} ~w[\lfloor u/l \rfloor].
\]\end{definition}
Let us note that, in the case $t = K  l_n$, $\tilde X_{l_n}(t)$ can be rewritten as
\[ \nonumber
\tilde X_{l_n}(t) =\sum_{k=0}^{K} e^{\omega_{l_n}(kl_n)} l_n w[k].
\]
This expression corresponds exactly to the original (log-normal) 
MRW expression introduced in Refs.
\cite{mrwepj,mrwpre}.
\section{Main results on MRW processes}
\label{MRWres}
\begin{theorem} 
({\bf The two subordinated MRW are the same processes})
\label{Tmrw}
If there exists $\epsilon > 0$ such that $\zeta_{1+\epsilon} > 1$, then
\[ \nonumber
X(t) =   \lim_{l\rightarrow 0^+} X_l(t)
\operatornamewithlimits{=}^{law}\lim_{l\rightarrow 0^+} B(M_l([0,t])) = B(M([0,t])) = X^{(s)}(t),
\]
where $B(t)$ (resp. $dW(u)$) is a Brownian motion (resp. Wiener noise) of variance 1 independent of $P(dt,dl)$.Moreover
\[ \nonumber
\<X(t)^2 \> = t.
\]
\end{theorem}
The proof of this theorem can be found in appendix \ref{amrw}.

\begin{theorem} {\bf ({\bf Main results on $X(t)$ and $X^{(e)}(t)$})} \\Let $\epsilon > 0$ such that $\zeta_{1+\epsilon} > 1$. 
Let $q > 0$. The following properties hold
\begin{itemize}
\item[(i)] $\zeta_{q} > 1 \Longrightarrow  \<|X(t)|^{2q}\>< +\infty$. 
\item[(ii)] $\<|X(t)|^{2q}\> < +\infty \Longrightarrow \zeta_q \ge 1$.
\item[(iii)] \[ \nonumber
\{X^{(e)}(t)\}_t ~\operatornamewithlimits{=}^{{law}} W_{\lambda}  \{X^{(e)}(t)\}_t,~~~\forall \; \lambda \in \; ]0,1[ \; \mbox{and} \; \; t \leq T \; ,
\]
with $W_{\lambda} = \lambda^{\frac{1}{2}}  e^{\frac{1}{2}\Omega_{\lambda}}$
where $\Omega_\lambda$ is an infinitely divisible 
random variable (independent of $\{X^{(e)}(t))\}_t$)  
which characteristic function is
\[ \nonumber
\< e^{iq\Omega_{\lambda}}\> = \lambda^{-\varphi(q)}.
\]
\item[(iv)] If  $\zeta_{q} = -\infty$, then $\<|X^{(e)}|(t)^{2q}\> =+\infty$ and otherwise
\[ \nonumber
\<|X^{(e)}(t)|^{2q}\> = \left(\frac{t}{T}\right)^{\zeta_{q}} \<|X^{(e)}(T)|^{2q}\>, ~~\forall t \le T.
\]
\item[(v)]  if $\<|X(t)|^{2q}\> < +\infty$ and $\exists \; \epsilon> 0$, 
$\zeta_{q+\epsilon} \neq -\infty$,
then
\[ \nonumber
\<|X(t)|^{2q}\>  ~\operatornamewithlimits{\sim}^{{t\rightarrow 0^+}} ~ \left(\frac{t}{T}\right)^{\zeta_{q}} \<|X(T)|^{2q}\>.
\]\end{itemize}
\end{theorem}
This theorem is a direct consequence of the theorems of section \ref{MRMres}.
Let us note that (iii) shows that $X^{(e)}$ is a ``continuous cascade'' 
as defined in (\ref{cascade}).

\begin{theorem} {\bf (Link between $\tilde X(t)$ and $X(t)$)}
If there exists $\epsilon > 0$ such that $\zeta_{2+\epsilon} > 1$ then
\[ \nonumber
\lim _{n \rightarrow +\infty} \{ X_{l_n} (t)\}_t \operatornamewithlimits{=}^{{law}} \{ X(t) \}_t.
\]
\end{theorem}
\begin{proof}
Since, we have proved (theorem \ref{Ttildem}) that as long as
there exists $\epsilon >0$ such that $\zeta_{2+\epsilon} > 1$ then $\tilde M_{l_n}(dt)$ converges towards $M(dt)$, one easily gets that 
the finite dimensional laws of 
$\{\tilde X_{l_n}(t)\}_t$ converge towards those of $\{X(t)\}_t$. 
Moreover, 
in the same way as at the end of appendix \ref{amrw}, 
using theorems of section \ref{MRMres}, one easily shows that the sequence
$\{\tilde M_{l_n}(dt)\}$ is tight and consequently $\{\tilde X_{l_n}(t)\}$ is also tight. This proves the theorem.
\end{proof}

\section{Examples of MRM or MRW - Connected approaches}
\label{examples}
Let us study, in this section, some examples of 
log-infinitely divisible MRM (or MRW).

Let us first remark that all along the paper we have
supposed that $q_c > 1$ (see Eqs.~(\ref{qc}) and (\ref{toto})).
That means that $\int_y^{+\infty} e^x x^{-2} \nu(dx) < +\infty$ 
and therefore, it is sufficient that 
\begin{equation}
\label{nucondition}
 \nu(dx) = O(x^{1-\epsilon} e^{-x}) \; \mbox{when} \; x \rightarrow +\infty
\end{equation}
Let us remark that this condition is notably 
satisfied for all Levy measures
which support is bounded to the right.

\begin{itemize}

\item {\em Lebesgue measure:} \\ 
The simplest case is when the Levy measure $\nu(dx)$
is identically zero. In that case the associated MRM is
trivially identical to the Lebesgue measure, i.e, $M(dt) = dt$,
and $\zeta_q = q$. \\

\item {\em Log-normal MRM:} \\ 
When the canonical measure $\nu$ 
attributes a finite mass at the origin, 
$\nu(dx) = \lambda^2 \delta(x) dx$ with $\lambda^2 > 0$,
it is easy to see from (\ref{LK}), that $\psi(p)$ is 
the cumulant generating function of a normal distribution:
$\psi(p) = pm +\lambda^2 p^2/2$. We thus have $q_c = +\infty$ and
the condition $\psi(1)=0$ implies the relationship $m=-\lambda^2/2$.
The log-normal $\zeta_q$ spectrum is thus parabola:
\begin{equation}
\label{lnzeta}
 \zeta_q^{\mbox{ln}} = q(1+\frac{\lambda^2}{2}) -\frac{\lambda^2}{2}q^2.
\end{equation}
In that case the associated MRW process is exactly the same
as the process defined in Refs. \cite{mrwepj,mrwpre}.
The Gaussian process $\omega_l(t)$ (definition \ref{defomega})
can be directly constructed by filtering a 1D white noise 
without any reference to 2D conical domains. 
This model is interesting because its
multifractal properties are described by
only two parameters, the integral scale $T$ 
and the so-called intermittency parameter
$\lambda^2$.
Moreover, many exact analytical expressions
can be obtained and notably the
value of the prefactor $\< M^{(e)}[0,T]^q \>$ in (\ref{mescaling}).
(see Refs. \cite{mrwphya,mrwpre}). \\

\item {\em Log-Poisson MRM:} \\ 
When there is a finite mass at some finite value $x_0= \ln(\delta)$,
of intensity $\lambda^2 = \gamma (\ln \delta)^2$: 
$\nu(dx) = \lambda^2 \delta(x-x_0)$.
The corresponding distribution is Poisson of scale parameter $\gamma$ and
intensity $\ln(\delta)$:
$\psi(p) = p\left(m-\sin(\ln(\delta))\right) -\gamma(1-\delta^p)$.
We have again $q_c = +\infty$ and 
the log-Poisson $\zeta_q$ spectrum is therefore:
\begin{equation}
\label{lpzeta}
  \zeta_q^{\mbox{lp}} = q m' + \gamma(1-\delta^q) 
\end{equation}
where $m'$ is such that $\psi(1) = 0$.
This situation corresponds to the model
proposed by She and L\'ev\^eque in
the field of turbulence \cite{sl}. \\

\item {\em Log-Poisson compound MRM:} \\
When the canonical measure $\nu(dx)$ is such that 
$\int \nu(dx) x^{-2} = C < +\infty$
(e.g., if $\nu$ is concentrated
away from the origin) 
it is easy to see that $F(dx) = \nu(dx)x^{-2}/C$ is a probability measure. 
In that case, 
\[
 \varphi(p) = im'p + C \int (e^{ipx}-1) F(dx)
\]
is exactly the cumulant generating function 
of a Poisson process with
scale $C$ and compound with the distribution $F$ \cite{Feller}.
Let us now consider a random variable $W$ such that
$\ln W$ is distributed according to $F(dx)$.
In this example,  if $q_c = \max_q \{ \< W^q \> < +\infty\} > 1$,
the log-Poisson compound MRM
has the following multifractal spectrum:
\begin{equation}
\label{lpczeta}
  \zeta_q^{\mbox{lpc}} = qm-C\left(\<W^q\>-1\right)
\end{equation}
The so-obtained MRM corresponds exactly to
Barral and Mandelbrot's ``product of cylindrical pulses'' in 
\cite{barral}. 
Let us note that these authors did not study the scaling
properties of their construction. They rather focused 
on the pathwise regularity properties.
They proved the validity
of the so-called ``multifractal formalism'' 
(see e.g. \cite{wtmm1,wtmm2,wtmm3,fp,jaff1,jaff2}) that relates
the function $\zeta_q$ to the singularity
spectrum $D(h)$ associated with (almost) all realisations
of the process by a Legendre transformation. \\

\item {\em Log-$\alpha$ stable MRM:} \\
Let us consider the case when $\nu(dx)$ corresponds to
a left-sided $\alpha$-stable density:
\[
  \nu(dx) = 
\left\{
\begin{array}{ll} 
C |x|^{1-\alpha} & \mbox{if}~ x \leq 0 \\
0 & \mbox{if}~ x > 0
\end{array}
\right. 
\] 
Where $C > 0$ and $0 < \alpha < 2$. 
In such a case, a direct computation shows that
$q_c = +\infty$: 
\begin{equation} 
\label{lszeta}
 \zeta_q^{\mbox{ls}} = qm-\sigma^{\alpha} |q|^{\alpha}
\end{equation}
Such laws have been used in the context of 
turbulence and geophysics \cite{scherlev}. \\

\item {\em Log-Gamma MRM :\\}
The family of Gamma distributions corresponds to 
Levy measures of the form $\nu(dx) = C \gamma^2 x e^{-\gamma x} dx$ for
$x \geq 0$. 
It is immediate to see that $q_c = \gamma$ for
this class of measures and thus one must have
$\gamma > 1$ in order to construct a MRM.
A direct computation leads to $\psi(q) = C \gamma^2 \ln(\frac{\gamma}{\gamma-q})$ and therefore, for $q < \gamma$,
\begin{equation}
  \zeta_q^{\mbox{lg}} = qm - C \gamma^2 \ln(\frac{\gamma}{\gamma-q})
\end{equation}

\end{itemize}
Many other families of $\zeta_q$ spectra can be obtained
for other choices of the Levy measure.

\section{Conclusion}
\label{conclusion}

In this paper we have proposed a
new construction of stationary stochastic measures and
random processes with stationary increments. We proved they have exact
multifractal scaling in the sense of (\ref{scaling}) (in which case it
satisfies the cascading rule (\ref{cascade}))
or asymptotic multifractal scaling properties in the sense
of (\ref{ascaling}). Apart from their multifractal properties, we have also
studied non degeneracy and conditions for finite moments.

There are many applications these processes can be used for. Actually, in a
previous work, we have already shown \cite{mrwepj} that the log-normal MRW
is a very good candidate for modeling financial data. However, multifractal
scaling are observed in many other fields ranging from turbulence to
network traffic or biomedical engineering. Some of these
possible applications of MRM and
MRW are discussed in \cite{premrw}.

It remains many open mathematical problems related to these processes. Some
of them are discussed in \cite{premrw},
notably the questions
related to the construction of stochastic integrals using fractional
Brownian motions instead of regular Brownian motions as in section
\ref{MRWdef}.
Another interesting problem concerns the study of
limit probability distributions associated with MRM for which
fery few features are known. Finally, it should be interesting
to generalize the results of  \cite{barral}
in order to link scaling properties and pathwise regularity
within a multifractal formalism.

\newpage 

\vskip 1cm
\begin{center}
  {\Large \bf Appendices}
\end{center}
\appendix
\section{Proof of lemma  \ref{lmchar} (characteristic function of $\omega_l(t)$)}
\label{Achar}
We are going to compute $Q_l(\vec{t}_{q} ,\vec{p}_{q} )$ using a recurrence on $q$.
If we group in the sum  $\sum_{k=1}^q p_k P(\A_{l}(t_k))$ the points wich are
not in the set
$\A_l(t_q)$, we get the random variable
\[ \nonumber
Y_{q} = \sum_{k=1}^{q} p_k P\left(\A_l(t_k) \backslash \A_l(t_q)\right).
\]
Moreover, the points that are in the set $\A_l(t_q)$ can be grouped using
the disjoint sets
$\B_k  = \A_l(t_k,t_q) \backslash \A_l(t_{k-1},t_q)$ (i.e., points belonging
to $\A_l(t_k)$ and not to $\A_l(t_{k-1})$). Thus, if we define
\[ \nonumber
X_{k,q} = P\left(\A_l(t_k,t_q) \backslash \A_l(t_{k-1},t_q)\right),
\]
(where  we used the notation $\A(t_0,t_m) = \A(t_m,t_0) = \emptyset$) then
one has
\[ \nonumber
\sum_{k=1}^q p_k P(\A_{l}(t_k)) = Y_q + \sum_{k=1}^{q} r_{k,q} X_{k,q},
\]
where the numbers $r_{q,k}$ are defined by
\[ \nonumber
r_{k,j} = \sum_{m=k}^j p_m.
\]
Moreover, since  all the $\{X_{k,q}\}_k$ and $Y_q$ are independent random
variables, one gets
\begin{equation}
\label{aa}
Q_l(\vec{t}_{q},\vec{p}_{q} )  = \<e^{iY_q}\> \prod_{k=1}^{q}
\<e^{i r_{k,q} X_{k,q}}\>
\end{equation}

The same type of arguments can be used to prove that
\[ \nonumber
\sum_{k=1}^{q-1} p_kP(\A_{l}(t_k)) = Y_q + \sum_{k=1}^{q} r_{k,q-1} X_{k,q},
\]
and
\begin{equation}
\label{ab}
Q_l(\vec{t}_{q-1},\vec{p}_{q-1})  =  \<e^{iY_q}\>
\prod_{k=1}^{q} \<e^{i r_{k,q-1} X_{k,q}}\>,
\end{equation}
where we used the convention $r_{k,j} = 0$ if $j< k$.
Merging (\ref{aa}) and (\ref{ab}) leads to
\begin{equation}
\label{ac}
Q_l(\vec{t}_{q},\vec{p}_{q}) = Q_l(\vec{t}_{q-1},\vec{p}_{q-1})  \prod_{k=1}^{q} \frac{\<e^{ir_{k,q}
X_{k,q}}\>}{\<e^{i r_{k,q-1} X_{k,q}}\>}.
\end{equation}
Since, using (\ref{meerde}), one gets
\[ \nonumber
\<e^{ip X_{k,q}}\> = \<e^{ipP\left(\A_l(t_k,t_q) \backslash
\A_l(t_{k-1},t_q)\right)}\> =
e^{\varphi(p)\left(\mu(\A_l(t_k,t_q)\backslash \A_l(t_{k-1},t_q))\right)}.
\]
However, since  $t_{k-1}\le t_k  \le t_q$, one has $\A_l(t_{k-1},t_q) \subset
\A_l(t_k,t_q)$, therefore
\[ \nonumber
\<e^{ip X_{k,q}}\> =
e^{\varphi(p)\left(\mu(\A_l(t_k,t_q)-\mu(\A_l(t_{k-1},t_q))\right)}.
\]
By inserting this last expression in (\ref{ac}), it follows
\begin{eqnarray}
\nonumber
Q_l(\vec{t}_{q},\vec{p}_{q})& = & Q_l(\vec{t}_{q-1},\vec{p}_{q-1})  \prod_{k=1}^{q}
\frac{e^{\varphi(r_{k,q})\left(\mu(\A_l(t_k,t_q))-\mu(\A_l(t_{k-1},t_q))\right)}}{e^{
\varphi(r_{k,q-1})
\left(\mu(\A_l(t_k,t_q))-\mu(\A_l(t_{k-1},t_q))\right)}}
\\
\nonumber
& = & Q_l(\vec{t}_{q-1},\vec{p}_{q-1})  \prod_{k=1}^{q}
e^{(\varphi(r_{k,q})-\varphi(r_{k,q-1}))(\mu(\A_l(t_k,t_q))-\mu(\A_l(t_{k-1},t_q)))}.
\end{eqnarray}
By iterating this last expression, one gets
\begin{eqnarray}
\nonumber
\ln Q_l(\vec{t}_{q},\vec{p}_{q}) & = & \sum_{j=1}^q \sum_{k=1}^j
(\varphi(r_{k,j})-\varphi(r_{k,j-1}))(\mu(\A_l(t_k,t_j))-\mu(\A_l(t_{k-1},t_j))) \\
\nonumber
 & = & \sum_{j=1}^q \sum_{k=1}^j (\varphi(r_{k,j})-\varphi(r_{k,j-1}))\mu(\A_l(t_k,t_j))   \\
\nonumber
& & - \sum_{j=1}^q  \sum_{k=1}^j (\varphi(r_{k,j})-\varphi(r_{k,j-1}))\mu(\A_l(t_{k-1},t_j)) \\
\nonumber
 & = & \sum_{j=1}^q \sum_{k=1}^j (\varphi(r_{k,j})-\varphi(r_{k,j-1}))\mu(\A_l(t_k,t_j))  \\
\nonumber
& &  -  \sum_{j=1}^q \sum_{k=0}^{j-1} (\varphi(r_{k+1,j})-\varphi(r_{k+1,j-1}))\mu(\A_l(t_{k},t_j))
\\
\nonumber
 & = & \sum_{j=2}^q \sum_{k=1}^{j-1}
(\varphi(r_{k,j})+\varphi(r_{k+1,j-1})-\varphi(r_{k,j-1})-\varphi(r_{k+1,j}))\mu(\A_l(t_k,t_j)) \\
\nonumber
& & + \sum_{j=1}^q \varphi(p_j) \mu(\A_l(t_j))-
\sum_{j=1}^q (\varphi(r_{1,j})-\varphi(r_{1,j-1}))\mu(\A_l(t_0,t_j)).
\end{eqnarray}
Since (i) $\varphi(0) = 0$ and (ii) by convention $\mu(\A_l(t_0,t_j)) = 0$,  one finally gets the lemma \ref{lmchar}
\qed

\section{Controlling the moments of $\sup_{u \in [0,t]} e^{\omega_l(u)}$}
In this appendix we show the following lemma
\begin{lemma}
\label{doob}
If $q$ is such that $\psi(q) \neq +\infty$ (i.e., $\zeta_q \neq -\infty$) then 
\[ \nonumber
\<\sup_{u \in [0,t]} e^{q\omega_l(u)} \> < +\infty
\]
\end{lemma}
\begin{proof}
By definition, one has $\omega_l(u) = P(\A_l(u))$.
We first consider $t$ small enough such that $\cap_{u\in [0,t]} \A_l(u)  = \A^{(i)}_l \neq \emptyset$.
Thus, for any $u \in [0,t]$, one can decompose $\A_l(u)$ into the three disjoint sets
\[ \nonumber
\A_l(u) = \A^{(i)}_l\cup \A^{(l)}_l(u) \cup \A^{(r)}_l(u),
\]
where $\A^{(l)}_l(u)$ (resp. $\A^{(r)}_l(u)$) corresponds to the part of $\A_l(u)$ which is on the
left (resp. right) of $\A^{(i)}_l(u)$. Thus
\[ \nonumber
\<\sup_{u \in [0,t]} e^{q\omega_l(u)} \>= \<e^{ qP(\A^{(i)}_l)} \>\<\sup_{u \in [0,t]} e^{ qP(\A^{(l)}_l(u))} \>
\<\sup_{u \in [0,t]} e^{ qP(\A^{(r)}_l(u))} \>.
\]
Since $P(\A^{(r)}_l(u))$ is a martingale, using Doob's
$L^p$ inequality for submartingales suprema \cite{Doob}, we get\[ \nonumber
\<\sup_{u \in [0,t]} e^{ qP(\A^{(r)}_l(u))} \> \le D_q \< e^{ qP(\A^{(r)}_l(t))} \>,
\]
where $D_q$ is a constant which depends only on $q$.
In the same way $P(\A^{(l)}_l(t-u))$ is a martingale, thus
\[ \nonumber
\<\sup_{u \in [0,t]} e^{ qP(\A^{(l)}_l(u))} \> \<\sup_{u \in [0,t]} e^{ qP(\A^{(l)}_l(t-u))} \> \le D_q \< e^{ qP(\A^{(r)}_l(0))} \>.
\]
This prooves the lemma.

In the case $t$ is large enough such that  $\cap_{u\in [0,t]} \A_l(u)  = \A^{(i)}_l \neq \emptyset$, we split the interval into $n$ equal intervals and we get
\[ \nonumber
\<\sup_{u \in [0,t]} e^{q\omega_l(u)} \> \le n\<\sup_{u \in [0,t/n]} e^{q\omega_l(u)} \>,
\]
and by choosing $n$ large enough we can use the same arguments as before.
\qed
\end{proof}

\section{Martingale properties of MRM}
Let us show the following lemma:
\begin{lemma}
\label{martin}Let $q>1$ such that $\psi(q) \neq +\infty$ (i.e., $\zeta_q \neq -\infty$).
For all fixed value of
$t$, the sequence $M_l([0,t])^q$ is a positive submartingale, i.e.,
$\forall \; l'<l$:
\[ \nonumber
 \< M_{l'}([0,t])^q || \F_l \> \geq M_l([0,t])^q.
\]Consequently, at fixed $t$, the sequence
$\< M_l([0,t])^q \>$ increases when $l$ decreases.\end{lemma}
\begin{proof}
Let $l'< l$, then since $\psi(1)=0$, 
\[ \nonumber
  \< M_{l'}([0,t]) || \F_l \>  =  \int_0^t \< e^{\omega_{l'}(t)} dt|| \F_l \> dt   = M_l([0,t]).
\]
$M_l([0,t])$ is therefore at positive martingale.
If $q > 1$, the submartingale property directly results
from Jensen inequality. \qed.
\end{proof}

\section{Proof of lemma \ref{LMRMmom} (Moments of positive orders of $M^{(e)}(dt)$)}
\label{Amom}
For the proof of this lemma, we proceed along the same line
as in Refs. \cite{barral,kp}. \\
{\bf Proof for (ii)}.\\
First, let us note that, since $M^{(e)} \neq 0$ and  $\<M^{(e)}([0,t])^q\> <+\infty$ then (using theorem \ref{TMRMescaling} which is a direct consequence of lemma \ref{lmexact}) one easily shows  that $\psi(q) \neq +\infty$, i.e., $\zeta_q \neq -\infty$. 
Using the superadditivity of $x^q$ for $q \geq 1$, one gets
\begin{eqnarray}
\nonumber
\<M^{(e)}([0,t])^q\> & = & \<(M^{(e)}([0,t/2])+M^{(e)}([t/2,t]))^q\> \\
\nonumber
& \ge & \<M^{(e)}([0,t/2])^q\>+\<M^{(e)}([t/2,t])^q\> \\
\nonumber
& \ge & 2\<M^{(e)}([0,t/2])^q\>.
\end{eqnarray}
Since $M^{(e)} \neq 0$ and $\zeta_q \neq -\infty$, using (\ref{mescaling}), 
we have
\[ \nonumber
\<M^{(e)}([0,t])^q\> \ge  2^{1-\zeta_q} \<M^{(e)}([0,t])^q\> \; ,
\]
and consequently $\zeta_q \ge 1$.\\
\vskip .2cm
\noindent
{\bf Proof for (i).} \\
Since $\zeta_q > 1$,  $\psi(q) \neq +\infty$ and thus 
(according to lemma \ref{doob}), one gets
\begin{eqnarray}
\nonumber
\<M^{(e)}_l([0,t])^q\> &= &\<(\int_0^t e^{\omega_l(u)} du )^q\> \\
\label{ccc}
&\le & \< \sup_{u \in [0,t]}  e^{q\omega_l(u)} \> t^q  < +\infty.
\end{eqnarray}
Let $m \in \NN$. Let us decompose $M^{(e)}_l$ as:
\[ \nonumber
M^{(e)}_l([0,T]) = M^{(0)}_l(T) + M^{(1)}_l(T),
\]
where
\begin{equation}
\label{bbb}
M^{(0)}_l(T) = \sum_{k=0}^{2^{m-1}-1} d_{2k},
\end{equation}
and
\[ \nonumber
M^{(1)}_l(T) = \sum_{k=0}^{2^{m-1}-1} d_{2k+1},
\]
with 
\[ \nonumber
d_k = M^{(e)}_l([kT2^{-m},(k+1)T2^{-m}]).
\]
Let us note that $M^{(0)}_l(T)$ and $M^{(1)}_l(T)$ are random variables 
identically distributed. Since $q \ge 1$, from Minkowski
inequality, one gets
\begin{equation}
\label{mink}
\<M^{(e)}_l([0,T])^q\> \le \left( \<M^{(0)}_l(T)^q\>^{\frac{1}{q}} + 
\<M^{(1)}_l(T)^q\>^{\frac{1}{q}} \right)^q = 2^q \<M^{(0)}_l(T)^q\>.
\end{equation}
Let $n \in \NN^*$ such that $n-1 < q \le n$. Thus using (\ref{bbb}) and
(\ref{mink}), we obtain:
\[ \nonumber
\<M^{(e)}_l([0,T])^q\>  \le 2^q \<(\sum_{k=0}^{2^{m-1}-1} d_{2k})^q\>.
\]
Thanks to the sub-additivity 
of the function $x^h$ (for $h = q/n \le 1$), one gets
\begin{eqnarray}
\label{eee}
\<M^{(e)}_l([0,T])^q\>  & \le & 2^q \<(\sum_{k=0}^{2^{m-1}-1} d_{2k}^{q/n})^n\>.
\end{eqnarray}
If we expand the last expression, 
the diagonal term gives simply $2^q 2^{m-1} \<d_0^q\>$. 
The non diagonal terms are of 
the form $C_{q,m} \<d_{2k_1}^{qs_1/n} \ldots d_{2k_n}^{qs_n/n}\>$ 
where $\sum_{i=1}^n s_i = n$ and $1 \le s_i < n$. 
Moreover, each $d_{2k}$ can be written as:
\begin{eqnarray}
\nonumber
d_{2k} &= &\int_{2kT2^{-m}}^{(2k+1)T2^{-m}} e^{\omega_l(u)} du , \\
\nonumber
& = & \int_{2kT2^{-m}}^{(2k+1)T2^{-m}} e^{\omega_{T2^{-m}} (u)} e^{\omega_{l,T2^{-m}}(u)} du,
\end{eqnarray}
where 
\[ \nonumber
\omega_{l,L}(u) = \omega_{l}(u)-\omega_{L}(u).
\]
Thus
\begin{equation}
\label{ddd}
d_{2k}  \ge  \inf_{v \in [0,T]} (e^{\omega_{T2^{-m}} (v)}) \int_{2kT2^{-m}}^{(2k+1)T2^{-m}} e^{\omega_{l,T2^{-m}}(u)} du,
\end{equation}
and
\[ \nonumber
d_{2k}  \le  \sup_{v \in [0,T]} (e^{\omega_{T2^{-m}} (v)}) \int_{2kT2^{-m}}^{(2k+1)T2^{-m}} e^{\omega_{l,T2^{-m}}(u)} du.
\]
This last expression can be seen as the product of 2 terms : the $\sup$ term and the integral term. They correspond to independent random variables. Moreover, since the $\sup$ term does not depend on $k$ and  for two different values of $k$  the integral terms are indepedant random variables, one finally gets for each non diagonal term
\begin{eqnarray}
\nonumber
\<\prod_{i=1}^r d_{2k_i}^{qs_i/n}\> & \le &\< \sup_{v \in [0,T]}
e^{q\omega_{T2^{-m}} (v)} \> 
\prod_{i=1}^r  \< \left(\int_{2k_iT2^{-m}}^{(2k_i+1)T2^{-m}} du ~e^{\omega_{l,T2^{-m}}(u)}\right)^{qs_i/n}\>
 \\
\nonumber
& \le & \< \sup_{v \in [0,T]} e^{q\omega_{T2^{-m}} (v)}\> 
\prod_{i=1}^r  \< \left(\int_{0}^{T2^{-m}} du ~e^{\omega_{l,T2^{-m}}(u)}\right)^{qs_i/n}\>.
\end{eqnarray}
From (\ref{ccc}), one knows that the $\sup$ term is bounded by
a finite positive constant $D$ which depends only on $q$, $m$ and $T$  
(and not on $l$).
Moreover, since $s_i < n$, one has $qs_i/n \le n-1$ and,
using H\"older inequality, the non-diagonal term can be bounded as:
\begin{eqnarray}
\nonumber
\<\prod_{i=1}^r d_{2k_i}^{qs_i/n}\>  & \le & 
D \< (\int_{0}^{T2^{-m}} du ~e^{\omega_{l,T2^{-m}}(u)})^{n-1}\>^{q
  \frac{\sum_{i=1}^{n}s_i}{n(n-1)}} \\
\nonumber
  & = &  D \< (\int_{0}^{T2^{-m}} du 
  ~e^{\omega_{l,T2^{-m}}(u)})^{n-1}\>^{\frac{q}{n-1}} 
\end{eqnarray}
On the other hand, from (\ref{ddd}), one gets 
\[ \nonumber
\< d_0^{n-1}\> \ge \<\inf_{v \in [0,T]} e^{(n-1)\omega_{T2^{-m}} (v)} \> \< (\int_{0}^{T2^{-m}} e^{\omega_{l,T2^{-m}}(u)} )^{n-1}\>.
\]Let us note that $\<\inf_{v \in [0,T]} e^{(n-1)\omega_{T2^{-m}} (v)} \> \neq 0$. Indeed, otherwise it would mean that 
{\em a.s.} there exists a sequence $\{v_p\}_p$ in $[0,T]$ 
such that $\lim_{p \rightarrow \infty} \omega_{T2^{-m}} (v_p) = -\infty$. This is impossible because $\omega_{T2^{-m}}(v)$ is cadlag.
Therefore, 
there exists a finite constant E which does not depend on $l$ such that
\[ \nonumber
\<\prod_{i=1}^r d_{2k_i}^{qn_i/n}\> \le E \< d_0^{n-1}\>^{\frac{q}{n-1}}
\]
Going back to (\ref{eee}), we finally proved 
that there exists a constant $C_{m,q}(T)$ (which does not depend on $l$) 
such that
\[ \nonumber
\<M^{(e)}_l([0,T])^q\>   \le  2^q 2^{m-1} \<d_0^q\> + C_{m,q}(T) 
\< d_0^{n-1}\>^{\frac{q}{n-1}}.
\]
Using the self-similarity of $M_l^{(e)}$ (lemma \ref{lmexact}), one gets for any $p$, $0<p \le q $ (and thus $\zeta_p \neq -\infty$) 
\[ \nonumber 
\< d_0^p\> = \<M^{(e)}_l([0,T2^{-m}])^p\> = 2^{-m\zeta_p} \<M^{(e)}_{2^ml}([0,T])^p\>.
\]
Thus
\begin{eqnarray}
\nonumber
\<M^{(e)}_l([0,T])^q\> & \le & 2^{q-1+m(1-\zeta_q)} \<M^{(e)}_{2^ml}([0,T])^q\> \\
\nonumber
& + & C_{m,q}(T)2^{-mq\frac{\zeta_{n-1}}{n-1}} 
\<M^{(e)}_{2^ml}([0,T])^{n-1}\>^{\frac{q}{n-1}}.\end{eqnarray}
Using lemma \ref{martin}, we know that, for $k \geq 1$, 
$\<M^{(e)}_{2^ml}([0,T])^{k}\> \le \<M^{(e)}_{l}([0,T])^{k}\>$ and
therefore $2^{ml}$ can be replaced by $l$ is the r.h.s. of 
the previous inequality.  
Since $\zeta_q > 1$, one can choose $m$ such that $q-1+m(1-\zeta_q)<0$, thus there exists a finite positive constant
$D_{m,q}(T)$ such that 
\begin{equation}
\label{sss}
\<M^{(e)}_l([0,T])^q\> \le  D_{m,q}(T)2^{-mq \frac{\zeta_{n-1}}{n-1}} 
\<M^{(e)}_{l}([0,T])^{n-1}\>^{\frac{q}{n-1}}
\end{equation}
Thus if $\sup_l \<M^{(e)}_{l}([0,T])^{n-1}\> < +\infty$ , then 
$\sup_l \<M^{(e)}_{l}([0,T])^{q}\> < +\infty$ and, since 
 $M_l^{(e)}([0,T])^q$
is a positive submartingale (see lemma \ref{martin}),  it converges and
$\<M^{(e)}([0,T])^q\> < +\infty$.

We are now ready to prove (i) by induction on $n$.
Let $q$ such that  $1< q \le  2$. In that case $n = 2$ and 
$\<M_l^{(e)}([0,T])^{n-1}\> =  T$. 
From the last assertion one deduces 
that $\sup_l \<M^{(e)}_{l}([0,T])^{q}\>  
< +\infty$ and $\<M^{(e)}([0,T])^q\> < +\infty$.
Thus it proves (i) for $q$ such that $1< q \le  2$. On the other hand, it also proves that, if $\zeta_2 > 1$
then $\sup_l \<M^{(e)}_{l}([0,T])^{2}\>  < +\infty$

Let us now suppose $2< q \le 3$, i.e., $n = 3$.
Since $\zeta_p$ is a concave function and $\zeta_1 = 1$ and $\zeta_q > 1$, 
one gets that  $\zeta_p >1$ for all $1< p\le q$, and in particular $\zeta_2 > 1$ and consequently$\sup_l \<M^{(e)}_{l}([0,T])^{2}\>  < +\infty$. 
Then again (\ref{sss}) gives that 
$\sup_l \<M^{(e)}_{l}([0,T])^{q}\>  < +\infty$
and (submartingale argument) 
$\<M^{(e)}([0,T])^q\> < +\infty$ which proves (i) 
for $q$ such that $2< q \le  3$.
By induction, applying the same arguments each time, one proves (i).
\qed

\section{Proof of lemma \ref{lmascaling} (Extension of the results on $M^{(e)}(dt)$ to  $M(dt)$)}
\label{Aascaling}We first consider the particular case where $f(l) = 0$ for $l>L$ (i.e., $g(l) = -f^{(e)}(l)$ in (\ref{fspec})). The so-obtained MRM will be referred to as $M^{(L)}$. \\
\vskip .2cm
\noindent{\bf Proofs for $M = M^{(L)}$} \\
Let $\A_l^{(e)}(t)$ be the domain in $\S^+$ associated with $M^{(e)}(dt)$, i.e., 
\[ \nonumber
\A_l^{(e)}(t) = \{(t',l'), ~ l'\ge l,~-f^{(e)}(l')/2 < t'-t\le f^{(e)}(l')/2\}.
\]
Let $\A_l(t)^{(L)}$ be the domain in $\S^+$ associated with $M^{(L)}(dt)$, i.e., 
\[ \nonumber
\A_l^{(L)}(t) = \{(t',l'), ~ L \ge l' \ge l,~-f^{(e)}(l')/2 < t'-t\le f^{(e)}(l')/2\}.
\]
It is clear that, if 
\[ \nonumber
\Delta_L(t) = \{(t',l'), ~ l'\ge L,~-f^{(e)}(l')/2 < t'-t\le f^{(e)}(l')/2\},
\]
then $\A_l^{(e)}(t) = \A_l^{(L)}(t) \cup \Delta_L(t)$ with $\A_l^{(L)}(t) \cap \Delta_L(t) = \emptyset$. Thus if
$\omega_l^{(L)}(t) = P(\A_l^{(L)}(t))$, $\omega_l^{(e)}(t) = P(\A_l^{(e)}(t))$ and $\delta_L(t) = P(\Delta_L(t))$,  one has
\[ \nonumber
\omega_l^{(e)}(t) = \omega_l^{(L)}(t)+\delta_L(t),
\]
where $\omega_l^{(L)}(t)$ and $\delta_L(t)$ are independent processes.
Thus
\[ \nonumber
M_l^{(e)}([0,t]) = \int_0^t e^{\omega_l^{(e)}(u)} du = \int_0^t e^{\omega_l^{(L)}(u)}e^{\delta_L(u)} du \; .
\]
Therefore,
\begin{equation}\label{bete1}
 M^{(L)}_l([0,t]) \inf_{v \in [0,t]} e^{\delta_L(v)} \le M_l^{(e)}([0,t]) \le  M^{(L)}_l([0,t]) \sup_{v \in [0,t]} e^{\delta_L(v)},
\end{equation}
and, taking the limit $l \rightarrow 0^+$,
\begin{equation}
\label{bete}
 M^{(L)}([0,t]) \inf_{v \in [0,t]} e^{\delta_L(v)} \le M^{(e)}([0,t]) \le  M^{(L)}([0,t]) \sup_{v \in [0,t]} e^{\delta_L(v)},
\end{equation}
Since $\delta_L(t)$ is a.s. right continuous and left-hand limited, one has 
\begin{equation}
\label{betemm}
\lim_{t\rightarrow 0^+} \inf_{v \in [0,t]} e^{\delta_L(v)} = \lim_{t\rightarrow 0^+} \sup_{v \in [0,t]} e^{\delta_L(v)} = 
e^{\delta_L(0)}.
\end{equation}
Moreover, since $ M^{(L)}_l([0,t])$ and $\delta_L(v)$ are independent processes, one gets from (\ref{bete})
\begin{eqnarray*}
\label{betem}
 \<M^{(L)}([0,t])^q\> \<\inf_{v \in [0,t]} e^{q\delta_L(v)}\> & \le & \<M^{(e)}([0,t])^q\> \\ & \le &  \<M^{(L)}([0,t])^q\> \<\sup_{v \in [0,t]} e^{q\delta_L(v)}\>.
\end{eqnarray*}
Let us note that, since $\delta_L(v)$ is cadlag, one has $\<\inf_{v \in [0,t]} e^{q\delta_L(v)}\> \neq 0$.
Then\begin{itemize}
\item[(i)] is an immediate consequence of (\ref{bete})
\item[(ii)] with $X=e^{\delta_L(0)}$ is an immediate consequence of (\ref{bete})
\item[(iii)] if $\zeta_q \neq -\infty$ (i.e., $\psi(q) \neq +\infty$) then $\<e^{q\delta_L(t)}\> < +\infty$ and, using the same argument as in lemma \ref{doob}, one gets that $\<\sup_{v \in [0,t]} e^{q\delta_L(v)}\> < +\infty$ and consequently (iii). \\
The case $\zeta_q = -\infty$ is a little trickier. Let us first note that, for $t$ small enough such that 
$\cap_{u\in[0,t]} \A_l(u) \neq \emptyset$, one has
\begin{equation}
\label{baba}
\<M_l^{(L)q}\> = \<(\int_0^t e^{\omega_l(u)}du)^q\> \ge C_t \< 
e^{qP(\cap_{u\in[0,t]} \A_l(u))} \> = +\infty.
\end{equation}
On the other hand, we know (lemma \ref{LMRMmom} (ii)) that 
$\<M^{(e)}([0,t])^q\> = +\infty$, so, in order to prove (iii) we would need to prove that
$\<M^{(L)}([0,t])^q\> = +\infty$. 
Since $\zeta_{1+\epsilon} > 1$ we know (lemma \ref{LMRMmom} (i)) 
that $\sup_l\<(M_l^{(e)}([0,t])^{1+\epsilon}\> < +\infty$.
Thus, using (\ref{bete1}), one gets $ \sup_l\<M_l^{(L)}([0,t])^{1+\epsilon}\>< +\infty$ and consequently $M_l^{(L)}$ is uniformly (in $l$) integrable
and $\< M^{(L)}([0,t]) || \F_l \> = M_l^{(L)}([0,t])$.
Thus, if we suppose $\<M^{(L)}([0,t])^q\> < +\infty$, one has 
\[ \nonumber
\<M^{(L)}([0,t])^q || \F_l \> \ge \<M^{(L)}([0,t]) || \F_l \>^q = M_l^{(L)}([0,t])^q.
\]
Consequently $\<M_l^{(L)}([0,t])^q\> \le \< M^{(L)}([0,t])^q\> < +\infty$ which contradicts (\ref{baba}). That completes the proof of (iii) in the case 
$\zeta_q = -\infty$.
\item[(iv)]  this is a direct consequence of (iii) along with  (\ref{betem}) and (\ref{betemm}). 
\end{itemize}\vskip .2cm
\noindent{\bf Proofs for any $M$} \\
So we just proved the theorem for $M = M^{(L)}$. Let now $g(l)$ be any function (satisfying the hypothesis (\ref{fspec})). Using exactly the same arguments as above (in which $M^{(e)}(dt)$ is replaced by $M(dt)$), and using that (i) and (ii) hold for $M^{(L)}$, one gets
(i) and (ii) for $M$. Moreover one gets
\begin{eqnarray}
\label{bibi}
 \<M^{(L)}([0,t])^q\> \<\inf_{v \in [0,t]} e^{q\delta'_L(v)}\> & \le & \<M([0,t])^q\> \\ \nonumber
& \le & \<M^{(L)}([0,t])^q\> \<\sup_{v \in [0,t]} e^{q\delta'_L(v)}\>,
\end{eqnarray}
where $\delta'_L$ is independent of $M^{(L)}$. Using that (iii) holds for $M^{(L)}$, (\ref{bibi}) implies that (iii) holds also for $M$. And (iv) is a direct consequence of (\ref{bibi}) and the fact that $\lim_{t\rightarrow 0^+} \inf_{v \in [0,t]} e^{\delta'_L(v)} = \lim_{t\rightarrow 0^+} \sup_{v \in [0,t]} e^{\delta'_L(v)} = 
e^{\delta'_L(0)}$.\qed

\section{Proof of theorem \ref{Ttildem} (link between $\tilde M$ and $M$)}
\label{a1}
One has
\begin{multline*} \nonumber
\< | M_{l_n}([0,t[) - \tilde M_{l_n}([0,t[) |^2\>  = \<  M_{l_n}([0,t[)^2 \> +  \<  \tilde M_{l_n}([0,t[)^2 \> \\
-2 \<  M_{l_n}([0,t[) \tilde M_{l_n}([0,t[) \>.
\end{multline*}
We are going to compute the limit of each term separately.
Since 
\[ \nonumber
\<  M_{l_n}([0,t[)^2 \>  =  \< \int_0^t  \int_0^t e^{\omega_{l_n}(u)+\omega_{l_n}(v)}dudv  \> 
 =  \int_0^t \int_0^t  e^{\varphi(2) \rho_{l_n}(u-v)}dudv .
\]
Thus
\[ \nonumber
\lim_{n \rightarrow +\infty} \<  M_{l_n}([0,t[)^2 \> = \int_0^t \int_0^t  e^{\varphi(2) \rho(u-v)}du dv  = 
\<  M([0,t[)^2 \> < +\infty.
\]
On the other hand 
\[ \nonumber
\tilde M_{l_n}([0,t[) = \sum_{k=0}^{t/l_n-1} e^{\omega_{l_n}(kl_n)} l_n 
 =  \int_0^t e^{\omega_{l_n}(\lfloor u/l_n \rfloor l_n)} du.
\]
Thus
\begin{eqnarray}
\nonumber
\<  \tilde M_{l_n}([0,t[)^2 \> & = & \< \sum_{k=0}^{t/l_n-1}  \sum_{k'=0}^{t/l_n-1}  
\int_0^t e^{\omega_{l_n}(\lfloor u/l_n \rfloor l_n) + \omega_{l_n}(\lfloor v/l_n \rfloor l_n)} du dv \> \\
\nonumber
& = & \int_0^t \int_0^t  e^{\varphi(2) \rho_{l_n}(\lfloor u/l_n\rfloor l_n - \lfloor v/l_n\rfloor l_n)}dudv .
\end{eqnarray}
Let
$\rho_l^{(M)}(t) = \rho_l(|t|-l)$ for $|t| \ge l$ and
$\rho_l^{(M)}(t) = \rho_l(0)$ for $|t| \le l$. Then, since $\rho_l(u)$ is a positive symetric decreasing function , one has $\forall u,v \ge 0$, $\rho_l(\lfloor u/l\rfloor l - \lfloor v/l\rfloor l) \le \rho_l^{(M)}(u-v)$. Moreover, since for any fixed $u$, one has $\lim_{l \rightarrow 0^+} \rho_l^{(M)}(u) = \rho(u)$, using the dominated convergence theorem, one gets
\[ \nonumber
\lim_{n \rightarrow +\infty} \<  \tilde M_{l_n}([0,t[)^2 \> = \int_0^t \int_0^t  e^{\varphi(2) \rho(u-v)}du dv =
\<  M([0,t[)^2 \>
\]
Now we just need to compute the limit of the cross term $\<  \tilde M_{l_n}([0,t[) M_{l_n}([0,t[) \>$.
In the exact same way as the last computation, we get
\[ \nonumber
\<  \tilde M_{l_n}([0,t[) M_{l_n}([0,t[) \> =  \int_0^t \int_0^t  e^{\varphi(2) \rho_{l_n}(\lfloor u/l_n\rfloor l_n - v)}dudv.
\]
Using the fact that $\forall u,v \ge 0$, $\rho_l(\lfloor u/l\rfloor l - v) \le \rho_l^{(M)}(u-v)$, we get
\[ \nonumber
\lim_{n \rightarrow +\infty} \<  \tilde M_{l_n}([0,t[) M_{l_n}([0,t[) \>  = 
 \int_0^t \int_0^t  e^{\varphi(2) \rho(u - v)}dudv = \<  M([0,t[)^2 \>
\]
\qed

\section{Proof of theorem \ref{Tmrw}}
\label{amrw}
{\bf Convergence of $X_l(t)$} \\
Let $0 \le t_1 \le t_2 \le \ldots \le t_n$. It easy to show that the $n$-dimensional random variable
$\{X_l(t_2)-X_l(t_1),\ldots, X_l(t_n)-X_l(t_{n-1})\}$ has the same law as
$\{M_l([t_1,t_2])w[1],\ldots, M_l([t_{n-1},t_n]) w[n-1])\}$, where $w[i]$ is a discrete Gaussion white noise of variance 1. Since $\zeta_{1+\epsilon} > 1$, according to theorem \ref{TMRMdeg} (i), we know that $M_l(dt)$converges and is not degenerated. Thus, the finite dimensional laws of the process $X_l(t)$ converge when $l\rightarrow 0^+$.

Moreover, let $q = 2(1+\epsilon')$, with $0 <\epsilon' <\epsilon$. One thus have $\zeta_{1+\epsilon'} > 1$.
\begin{eqnarray}
\nonumber
\< |X_l(t_2) - X_l(t_1)|^q\> & = & \< |\int_{t_1}^{t_2} e^{\omega_l(u)/2}dW(u)|^q \> \\
\nonumber
& = & \< M_l([t_1,t_2]) ^{q/2} \> \< |w|^q\>.
\end{eqnarray}
Let us notice that $\< |w|^q\>$ is finite and does not depend on $l$ and, using theorem \ref{TMRMmom} (ii) (along with the dominated convergence theorem), one has
\[ \nonumber
\lim_{l \rightarrow 0^+} \<M_l([t-1,t_2])^{1+\epsilon'}\>  = \<M([t-1,t_2])^{1+\epsilon'}\>  < +\infty,
\]
and, there exists a constant $C$ such that 
\[ \nonumber 
\<M([t-1,t_2])^{1+\epsilon'}\>  \le C|t_2-t_1|^{\zeta_{1+\epsilon'}},~~|t_2-t_1| \le 1.
\]
Thus, there exists a constant $D > 0$ such that 
\[ \nonumber
\< |X_l(t_2) - X_l(t_1)|^q\> \le D  |t_2-t_1|^{\zeta_{1+\epsilon'}},~~|t_2-t_1| \le 1.
\]
Since $\zeta_{1+\epsilon'} > 1$ this proves that $X_l$ is tight. Along with the convergence of the finite dimensional laws, it proves that $X_l$ converges when $l \rightarrow 0^+$.
\\
\\
{\bf Convergence towards B(M([0,t]))} \\
Since $\{M_l([t_1,t_2])w[1],\ldots, M_l([t_{n-1},t_n]) w[n-1])\}$ has the same law as \\
$\{B(M_l([t_1,t_2]))\ldots, B(M_l([t_{n-1},t_n]))\}$, we just need to prove that $B(M_l([0,t])) \rightarrow B(M[0,t])$. Using the same kind of arguments as above, one can show that $B(M_l([0,t]))$ is tight and that its finite dimensional laws converge to those of $B(M([0,t]))$. \qed

\end{document}